# Experimental study of the effect of phase mismatch on a CW-pumped cascaded quadratic nonlinear frequency comb

**Ville Ulvila[1] and Markku Vainio[2,3]**

[1] VTT Technical Research Centre of Finland Limited, Espoo, Finland
[2] Department of Chemistry, University of Helsinki, Helsinki, Finland
[3] Photonics Laboratory, Physics Unit, Tampere University, Tampere, Finland

E-mail: markku.vainio@helsinki.fi

**Abstract**

Optical frequency comb generation by continuous-wave-pumped cascaded quadratic nonlinearities (CQN) appears a promising alternative to well-established modelocked lasers, especially if access to the mid-infrared spectral region is needed. We report an experimental investigation of spectral properties of a CQN frequency comb, which is based on second-harmonic generation (SHG) and is pumped internally by a continuous-wave optical parametric oscillator. Our study focuses on SHG phase mismatch, which has drawn little attention in the previously reported CQN frequency comb research. The main observation of our study is that it is possible to improve spectral purity of the CQN frequency comb by adjusting the phase mismatch. In addition to the CQN process that generates a frequency comb in the near-infrared region, our experimental setup involves several other nonlinear processes. These processes lead to a composite comb, which extends from visible to mid-infrared.

## 1. Introduction

Among several other applications, optical frequency combs (OFCs) are excellent light sources for molecular spectroscopy and trace gas detection [1-3]. Compared to single-frequency lasers, OFCs can improve both the spectral coverage and resolution of spectroscopic measurements. On the other hand, certain OFC spectroscopy methods can yield sub-microsecond temporal resolution, for example to study chemical reactions in real time [4]. Both fundamental and applied molecular spectroscopy benefit from having a light source in the mid-infrared region, which accommodates the strong fundamental vibrational transitions of most molecules. The atmospheric windows in the 3 to 5 µm and 8 to 11 µm wavelength regions are particularly interesting for many applications, such as trace gas detection [1].

Despite the continuing progress in the development of quantum-cascade and interband-cascade laser technologies, nonlinear optics remains an important approach to produce coherent light in the mid-infrared region, especially if wide spectral coverage is required. In this paper, we investigate the cascaded quadratic nonlinear (CQN) frequency comb, which is a relatively new class of optical frequency combs [5-8]. The CQN comb generation is based on nonlinear phase shift that results from two consecutive second order $\chi^{(2)}$ nonlinear processes, typically second harmonic generation (SHG) followed by a back-conversion process (parametric down-conversion). We use the term CQN comb to differentiate the method from several other frequency-comb generation methods based on quadratic $\chi^{(2)}$ nonlinear processes, such as synchronously-pumped optical parametric oscillators (SP-OPOs) [3, 9-12] and electro-optic combs [13-16]. It is worth emphasizing that the CQN method generates the frequency comb from a continuous-wave (CW) pump field, while *e.g.* the SP-OPO requires a femtosecond OFC as an input.

One way to understand the principle of CQN comb generation is to examine the two successive nonlinear processes in frequency domain, see Fig. 1 [7, 17, 18]. The CQN comb generator has a $\chi^{(2)}$ nonlinear crystal placed inside an optical resonator and the resonator is pumped by a CW laser source at fundamental frequency $\omega_f$. The fundamental field builds up in the resonator, leading to SHG. Once the SH power at frequency $\omega_{sh} = 2\omega_f$ grows strong enough in the crystal, it can start an efficient back-conversion process, which can be understood as a doubly resonant OPO (DRO) pumped by the SH field. This back-conversion process can transfer energy not only to the originally oscillating cavity mode, but also to nearby cavity modes that fall within the parametric gain bandwidth. As these new frequencies are further up-converted (via SHG and sum-frequency generation SFG) and back-converted, a large number of resonant modes appears – see Fig. 1a and references [18, 19] for simplified schematic illustrations. If the resonant modes are mutually phase-locked by cross coupling of the fundamental and up-converted spectra, an equidistant frequency comb is produced at both spectral regions. The comb mode spacing is thus coarsely set by the dispersion-dependent free spectral range (FSR) of the cavity, but can be equalized by the nonlinear interactions. We have experimentally confirmed the comb uniformity in our previous work [19].

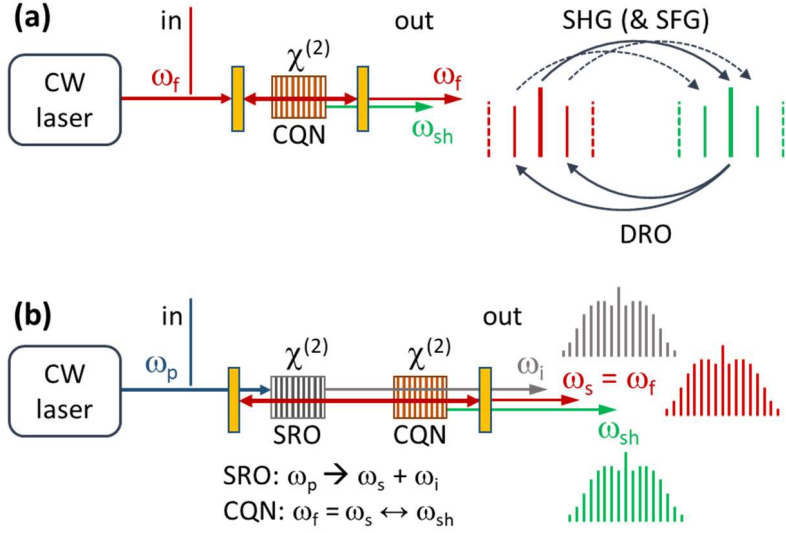

**Fig. 1.** (a) CQN comb generation with direct pumping of the $\chi^{(2)}$ crystal placed inside a cavity that resonates fundamental frequency $\omega_f$. (b) Internally pumped CQN comb generator. The resonant signal wave ($\omega_s$) of CW-pumped SRO acts as the fundamental wave ($\omega_f$) for CQN. The mid-infrared comb is centred at idler frequency, $\omega_i = \omega_p - \omega_s$.

A process similar to that outlined in Fig. 1a can also be pumped internally, by producing the fundamental wave within the same optical cavity that comprises the CQN crystal. An efficient way to do this is to use a singly-resonant CW OPO (SRO), which is schematically shown in Fig. 1b. This scheme was used in the work reported in this article, as well as in our first CQN comb demonstration in 2013 [5]. An advantage of the internally pumped system is that the SRO process automatically adjusts itself such that its signal wave ($\omega_s = \omega_f$) resonates in the cavity and constantly seeds the CQN process. In contrast, the external pumping (Fig. 1a) usually requires active locking of the pump laser frequency to a cavity resonance mode. Another advantage of the setup of Fig. 1b is that the SRO inherently and with high efficiency converts the CQN comb to the mid-infrared region [6, 19]. In our earlier work, we have experimentally shown that the mode spacing and offset frequency of the mid-infrared comb can be controlled independently, which is a prerequisite for full stabilization of the comb [6].

In addition to the schemes shown in Fig. 1, other implementations of CQN comb generation have been investigated both experimentally and theoretically during the past few years [20-22]. Much progress has been made in the development of theory and simulation tools to better understand the CQN comb process [8, 21-23]. These models make it possible to describe the comb sideband generation in terms of modulation instability (MI). In particular, the role of group velocity mismatch (temporal walk-off) has been examined in detail [8, 22, 23].

Most of the experimental work on CQN combs has so far used free-space optical cavities with bulk crystals (periodically poled lithium niobate, PPLN) but first steps towards miniaturized monolithic devices have been taken using waveguide resonators [24, 25] and whispering-gallery-mode microresonators [26, 27]. Impressive results in OFC generation by on-chip lithium niobate microresonators have already been reported, although the on-chip comb generation was attributed to third-order $\chi^{(3)}$ nonlinear processes [28, 29]. In general, the Kerr frequency combs based on $\chi^{(3)}$ nonlinear microresonators have attracted much more attention than the CQN combs, and various platforms have been developed [30, 31]. Interestingly, these two cases – the CQN comb and the Kerr comb – are in many ways analogous, because the cascaded second-order processes mimic third-order nonlinearities. This analogy hence provides another way to understand CQN comb formation. For example, even a slightly phase-mismatched SHG process may lead to a substantial nonlinear phase shift in the fundamental beam, thus producing an effective third-order nonlinearity that can be described by an effective nonlinear refractive index $n_2^{\text{eff}}$, and that manifests itself via effects similar to those caused by the Kerr nonlinearity [32, 33].

Especially the spatial and temporal properties of cascaded quadratic nonlinear effects were extensively studied in the 1990's, when spatial solitons and pulse compression by CQN were demonstrated [32]. Laser mode locking has been studied using, *e.g.*, CQN lensing, which mimics Kerr lensing [34, 35]. In laser mode locking, the nonlinear crystal is placed inside the laser cavity, which renders the case quite different from a CQN comb generator: As will be briefly discussed in this article, the lack of energy reservoir (laser gain material) is likely to complicate formation of temporal cavity solitons, *i.e.* short pulses that propagate in the cavity without changes in their energy or temporal shape. This is a topic of high practical importance. For instance, with Kerr frequency combs, access into the cavity soliton regime is a prerequisite for stable low-noise operation of the comb [36-38].

Despite the many similarities of the two approaches, frequency comb generation by CQN has potential advantages compared to Kerr combs. One obvious advantage is the easy access to the mid-infrared region, even with near-infrared laser pumping [6]. As the effective nonlinear refractive index due to CQN can be orders of magnitude higher than a material's inherent nonlinear refractive index $n_2$, the CQN method leads to high efficiency with a potential for low threshold. As an example, a CW-pumped CQN threshold as low as 2 mW has been recently observed using a lithium niobate microresonator [26]. The high efficiency also allows OFC generation with a free-space resonator that has a modest finesse and small FSR. This makes it possible to have a comb mode spacing of the order of 100 MHz, which is useful in many applications, such as high-resolution gas-phase spectroscopy. Another potential advantage of the CQN method is that, by adjusting the phase mismatch

$$\Delta k = k_{sh} - 2k_f - 2\pi/\Lambda \qquad (1)$$

of the SHG process[*], both the magnitude and sign of $n_2^{eff}$ can be controlled [5, 32, 33]. This makes it possible to compensate for both anomalous and normal group velocity dispersion, thus potentially allowing cavity soliton formation in both dispersion regimes.

The frequency-domain properties of CQN processes have so far drawn attention mainly in the context of supercontinuum generation by femtosecond-pumped CQNs [39-42]. An interesting exception is an early work on CW-pumped CQN wavelength conversion that mimics four-wave mixing (FWM) [43], which is one of the main third-order nonlinear processes behind Kerr comb generation. As the research of CW-pumped CQN resonators has started only quite recently, many details of CQN comb generation remain unexplored. For example, the role of phase mismatch $\Delta k$ has been mostly ignored in the previous research, although it is well known that CQN phenomena generally depend on $\Delta k$. While some of the earlier experimental and theoretical works have pointed out the importance of $\Delta k$ [5, 18, 44-46], detailed understanding of the effect of this parameter on CQN comb formation is missing.

Our experimental work reported in this paper focuses on the changes of spectral properties of a CQN comb as the phase mismatch is varied. Although we observe rather small effects in the CQN comb envelope spectrum due to $\Delta k$, our measurements reveal interesting details that are potentially useful, for instance, in view of generating soliton-mode-locked low-phase-noise CQN combs with high conversion efficiency. Our study also serves as a benchmark for future theoretical work, which is needed to better understand and quantify the role of $\Delta k$ and other parameters in CQN comb optimization.

## 2. Experimental setup

The experimental setup is shown schematically in Fig. 2. The Appendix contains more details about component specifications and other experimental details, including the group-delay dispersion (GDD) and reflectance spectra of the cavity mirrors. The $\chi^{(2)}$ nonlinear crystal responsible for the CQN action is placed inside a travelling-wave bow-tie resonator, which comprises mirrors that reflect the fundamental wave ($\omega_f$) but transmit the second-harmonic wave ($\omega_{sh}$). We produce the fundamental wave internally in the resonator by another nonlinear optical process, CW SRO. The CW SRO crystal is pumped at 1064 nm by a high-power narrow-linewidth CW laser system. The idler beam of the CW SRO is coupled out immediately after the CW SRO crystal, while the signal field feeds the CQN process. Cavity finesse at the resonant signal wavelength is about 300, as estimated from the calculated cavity losses. Approximately 2 to 2.5 W of pump power at 1064 nm is needed to start the CW SRO. As soon as the SRO oscillation threshold is exceeded and the fundamental wave is generated, we observe the onset of the CQN process. Our experimental setup does not allow reliable determination of the CQN comb threshold, since the CW SRO tends to be unstable when operated very close to its threshold. However, based on a previously reported experimental work of Ricciardi *et al.* [7], the CQN comb threshold in a typical free-space cavity can be estimated to be of the order of 100 mW, which in our intracavity-pumped system is reached almost immediately after exceeding the SRO threshold.

---

[*] Here, $k_f = \omega_f n_f / c_0 = 2\pi n_f / \lambda_f$ and $k_{sh} = \omega_{sh} n_{sh} / c_0 = 2\pi n_{sh} / \lambda_{sh}$ are the wave vectors of collinearly propagating fundamental and second harmonic waves, respectively. Symbols $n$ and $\lambda$ denote the respective refractive indices and wavelengths, and $c_0$ is the speed of light in vacuum. We have assumed a quasi-phase-matched nonlinear crystal that has a grating period of $\Lambda$. Although we limit our discussion to a CQN process that consists of SHG/SFG followed by down-conversion, a nonlinear phase shift can also be acquired with other cascaded processes.

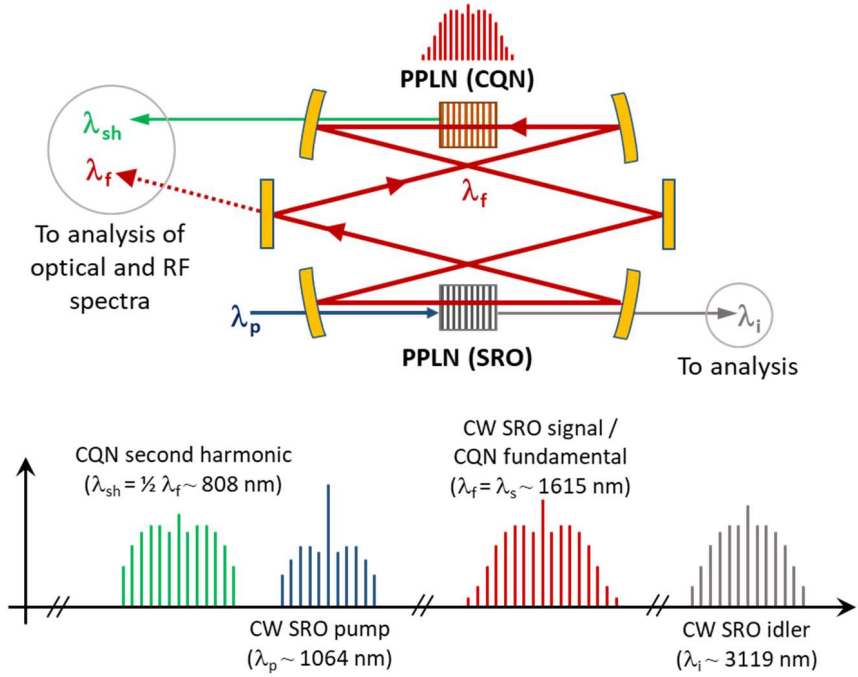

**Fig. 2.** Schematic illustration of the experimental setup, as well as the main wavelength regions of interest. The resonator is symmetric and consists of 2 plane mirrors and 4 concave mirrors, all of which are higly reflective in the fundamental wavelength range $\lambda_f$.

Both crystals (PPLN) are 50-mm long 5 % MgO-doped congruent LiNbO$_3$ crystals and have the poling periods designed such that all of the interacting waves have extraordinary polarizations. The main nonlinear processes that take place in the two crystals are:

- CW SRO crystal: 1064 nm (pump) → 1615 nm (signal) + 3119 nm (idler);
- CQN crystal: 1615 nm (fundamental) ↔ 807.5 nm (second harmonic + sum frequencies).

Spectral broadening and comb generation is observed at all these wavelengths [6, 17]. Most importantly, the CW SRO inherently converts the CQN comb to the mid-infrared region, where we can extract watt-level output power. Using the operating parameters that produce the best spectral quality with the current setup, we measure up to 1.5 W of mid-infrared output power after several dichroic mirrors and with the maximum CW SRO pump power of 19 W. This corresponds to an estimated mid-infrared power of > 2 W at the CW SRO crystal output. (The pump depletion in this case is 40 %, as discussed in more detail in Section 3).

The measurement instrumentation is described in detail in [17, 19]. In brief, the mid-infrared spectrum was monitored using a Fourier-transform infrared spectrometer (EXFO WA-1500/IR-89 + EXFO WA-650). The signal (fundamental) and second-harmonic beams – at 1615 mn and 807.5 nm, respectively – were split to an optical spectrum analyser (OSA; Ando AQ-6315E, 350–1750 nm) and fast photodiodes in order to characterize the optical and radiofrequency (RF) spectra of the respective combs.

Figure 3a shows the envelope spectrum of a composite comb that results from the various nonlinear mixing processes in the two crystals. An example of the mid-infrared envelope spectrum is plotted in Fig. 3b. The comb mode spacing is about 209 MHz, which is determined by the cavity FSR. Such a small mode spacing cannot be resolved in the optical spectra of Fig. 3, but we can observe and characterize the comb structure at high precision using, *e.g.*, the radiofrequency spectra.

The centre wavelengths of the different parts of the composite comb can be varied by changing the phase-matching conditions in both CW SRO and CQN crystals, in practice by adjusting the crystal poling periods and temperatures. As an example, we have demonstrated tuning of the mid-infrared comb wavelength from 3 to 3.4 µm [6]. Here, we use only the abovementioned single operating point and focus on the properties of CQN comb generation as a function of SHG phase mismatch $\Delta k$. In order to isolate the effect of $\Delta k$, the fundamental wavelength was fixed to 1615 nm and $\Delta k$ was varied by changing primarily the CQN crystal poling period $\Lambda$. In practice, this was done by translating the crystal, which has multiple poling periods in parallel; see Appendix for details. The operating wavelength was arbitrarily chosen within the flat region of both reflectivity and GDD of the cavity mirrors, see Fig. A1 of the Appendix. Because the CQN crystal has only discrete values of $\Lambda$, we also changed the crystal temperature in order to vary $\Delta k$ in finer steps. The value of $\Delta k$ at each measurement point was calculated using the Sellmeier equations for 5 % MgO-doped congruent LiNbO$_3$ [47]. The maximum temperature change between the measurement points reported here was 40 °C, resulting in just ~ 1 % change in the group-velocity dispersion caused by the CQN crystal.

The main experimental observations of the CQN comb spectral properties as a function of $\Delta k$ are presented in the following section. In addition to the 1615 nm CQN setup presented here, we have performed similar experiments with another setup, which was designed to operate at a fundamental (signal) wavelength of 2036 nm by using another set of cavity mirrors and nonlinear crystals [19]. This corresponds to operation near the PPLN zero-dispersion wavelength, as well as near the degeneracy wavelength of the 1064-nm pumped CW SRO. Despite these differences, the observed characteristics of the 1615 nm and 2036 nm systems are qualitatively similar.

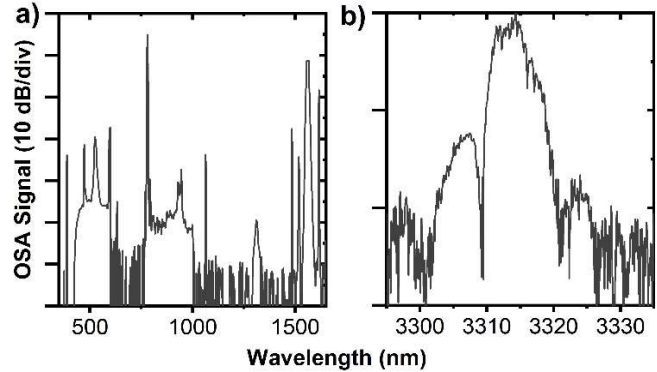

**Fig. 3.** The composite comb spectrum measured in two parts with two different optical spectrum analysers. (a) The strongest features are the fundamental and second-harmonic spectra. Higher harmonics extend to 390 nm, which is the shortest wavelength measurable with the optical spectrum analyser. (b) Envelope spectrum of the mid-infrared comb.

## 3. Results

### 3.1 Optical envelope spectra

Fig. 4 shows examples of optical spectra measured with different values of SHG phase mismatch $\Delta k$. The spectra are shown for the fundamental frequency range around $\omega_f$, as well as for the second harmonic frequency range around $\omega_{sh}$. (For convenience, we refer to the up-converted spectrum as "second harmonic (SH) spectrum", although it also contains components due to sum-frequency generation). The spectral range of the fundamental spectrum shown in Fig. 4 corresponds to the region of high reflectivity of the cavity mirrors. The mid-infrared idler spectra are not shown, as they remain essentially unchanged and are limited by the SRO phase-matching bandwidth, which is independent of the CQN crystal. As an example, Fig. 3b shows a typical idler spectrum.

The fundamental spectra have two distinct features regardless of $\Delta k$: a broad central peak and a number of side peaks. The central peak appears as soon as the SRO exceeds the threshold and starts to seed the CQN process at 1615 nm. The width of this peak depends on the phase mismatch (compare e.g. Figs. 4a and 4b), but also on the pump power – an example of the increasing width of the centre peak as a function of pump power is shown in Fig. A3 of the Appendix.

As the pump power is increased, the additional side peaks appear on both sides of the seed frequency. These additional peaks can be identified as signal and idler peaks of the DRO process pumped by the second harmonic field (Fig. 1). (The effective third-order nonlinearity model would ascribe these peaks to effective degenerate FWM arising from the CQN process). This interpretation of the side peaks is confirmed by the calculated parametric gain curves that are plotted with dashed lines on top of the measured fundamental spectra. These gain curves were calculated using a small-signal gain model of parametric amplification [48] and assuming a single-frequency second-harmonic beam at the frequency that corresponds to the strongest SH feature in the measured spectrum. (In reality, the DRO process is pumped by several SH frequencies, potentially leading to multiple sets of side peaks with different separations. For clarity, we have included the calculated curves for only one SH "pump frequency" in each plot, explaining the strongest side peaks.) We did not attempt to determine the absolute value of the gain or the effects of mirror or crystal coatings, so the gain spectra do not predict the relative heights of the side peaks but just their positions in frequency domain. It is useful to note that these DRO processes are generally phase mismatched, which explains the large number of side peaks in the gain spectra. In the more usual case of phase-matched DRO process, only two peaks – one signal and one idler – would appear in the spectrum.

Although the simple DRO model can qualitatively explain the main features of the spectra, it is clear that a detailed description would require a rigorous numerical model that includes also other nonlinear effects. For example, the broadening of the central peak is likely to involve MI dynamics, which for the $\Delta k$ values and operating conditions studied here typically produces sidebands a few hundred GHz away from the centre [44]. It is also worth emphasizing that formation of a phase-locked comb spectrum requires mutual coupling of the different resonator modes, which generally requires a cascade of two processes, such as SHG/SFG + back-conversion, as illustrated in Fig. A6 of the Appendix. This is analogous to mutual phase locking (injection locking) that stabilizes a degenerate synchronously-pumped OPO [49]. Within the framework of effective $\chi^{(3)}$ nonlinearity, this is analogous to self-injection locking of a Kerr comb [50].

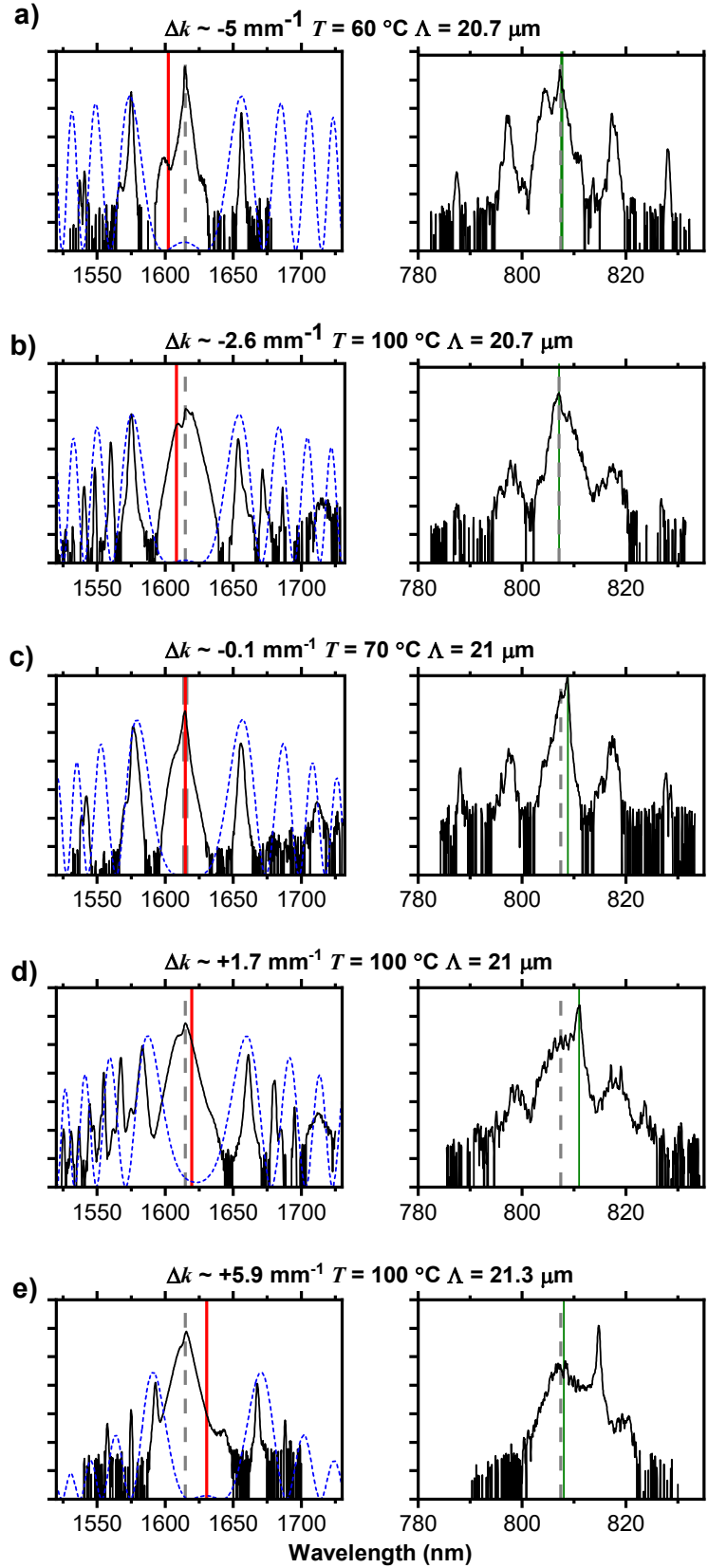

**Fig. 4.** Optical envelope spectra in the fundamental (left) and second-harmonic (right) spectral regions and with different values of $\Delta k$. The respective values of CQN crystal poling period and temperature are also indicated. Wavelengths of the CW seed (1615 nm) and its SH (807.5 nm) are denoted by dashed vertical lines. The red vertical line shows the wavelength for which $\Delta k = 0$. The green vertical line

indicates the SH wavelength used to simulate the DRO gain curve, which is outlined by the blue dashed curve and explained in the text. One division of the vertical axis corresponds to 10 dB.

## 3.2 Radiofrequency spectra

While the measurements of optical spectra clearly show spectral broadening and give interesting information of the nonlinear processes involved, they do not prove OFC generation. The existence of spectral features separated by one FSR of ~ 209 MHz can be confirmed with a high-resolution optical spectrum analyser [5], which is however insufficient to determine whether the comb is uniform. A more rigorous characterization of the comb can be done by measuring the radiofrequency spectrum, which comprises the intermode beat signals of the comb lines [6, 19]. We have previously carried out RF measurements to verify the CQN equidistance within 1 Hz, corresponding to a fractional precision of about $5 \times 10^{-15}$ and being limited by the resolution of our RF measurement instrument [19]. Moreover, the RF measurements give useful information about the comb phase noise and spectral quality of the comb.

Fig. 5 shows the measured RF spectra of the fundamental and second-harmonic combs for the same operating points that are shown in Fig. 4. In most cases, the beat note signal is rather broad, having a –3 dB linewidth of approximately 1 MHz. A particularly illustrative example of this is shown in Fig. 6a, which shows a broad beat signal that clearly consists of more than one overlapping beat notes. This behaviour is similar to what has been reported for $\chi^{(3)}$ Kerr combs and can be ascribed to slightly different mode spacings of the subcombs that corresponds to separate peaks in the optical envelope spectra [51]. So, although the comb spacing is uniform within a single subcomb [19], the mode spacings of different subcombs generally differ from each other. These mode spacings can be controlled by slightly varying the cavity length using a piezoelectric actuator, which is attached to one of the cavity mirrors. Even a small change in the cavity length due to a thermal drift *etc*. can change the intermode beat pattern [17]. This is exemplified in Fig. 6b, which shows the evolution of the RF spectrum when it is recorded over several seconds.

Another intriguing detail can be found when the subcombs start to overlap, but do not yet fully merge. In this case, the RF spectrum includes additional peaks, which is a signature of different offset frequencies of the subcombs [50, 52]. The additional peaks are notably stable and typically appear at $\frac{1}{2}\Delta\omega$, where $\Delta\omega$ ~ FSR refers to the comb mode spacing (Fig. 7). This corresponds to interleaved combs with a $\frac{1}{2}\Delta\omega$ difference in their offsets. Interestingly, Del'Haye *et al*. have found that for a microresonator Kerr comb the interleaved comb corresponds to near-zero detuning of the pump laser from the respective cavity mode [52]. Although we are currently unable to measure the cavity detuning, it is expected to be zero or close to zero owing to the self-adaptive CW-SRO signal wave that pumps the CQN comb.

In addition to better dispersion control of the resonator, the subcombs can be synchronized by external modulation, for example by seeding the resonator with a pump beam that is modulated at a frequency that matches the cavity FSR. The seeding technique has been demonstrated with both Kerr combs [51] and CQN combs [19].

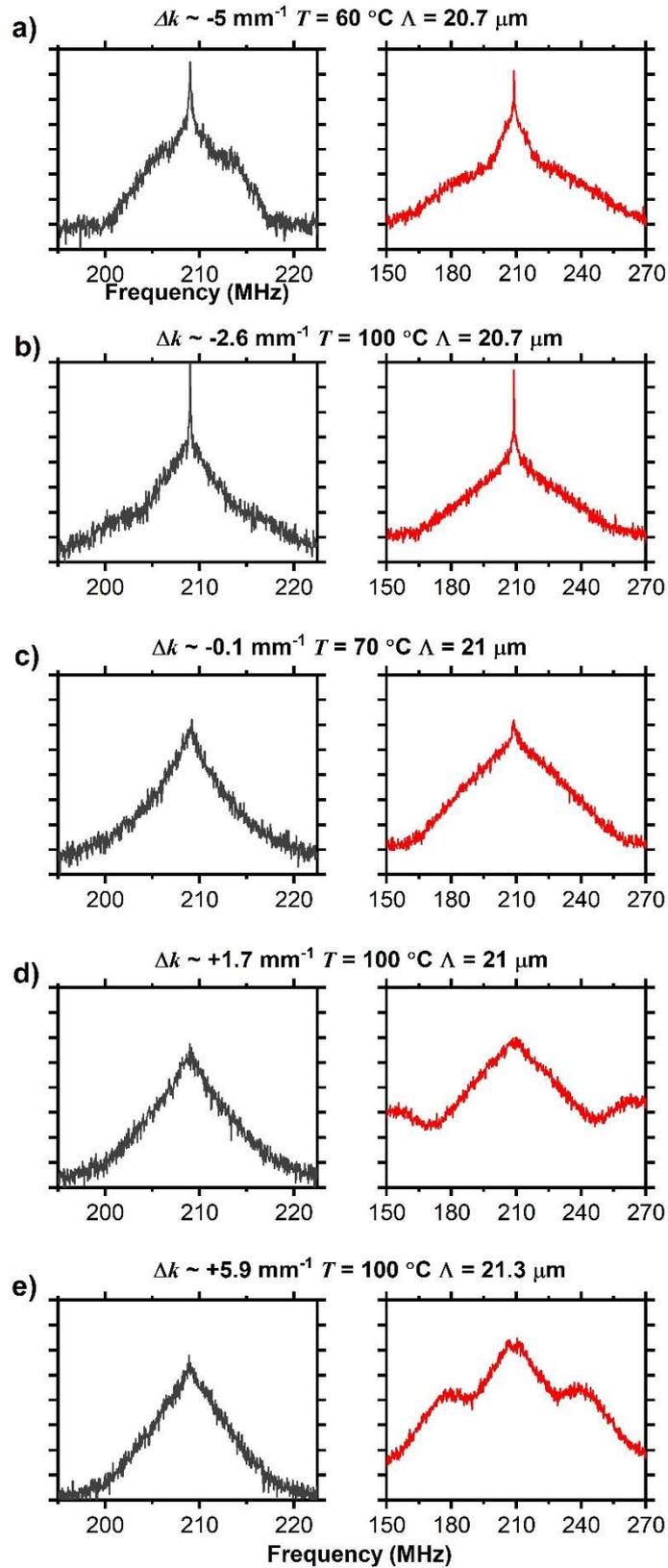

Fig. 5. Intermode beat notes (RF spectra) measured from the fundamental (left) and second-harmonic (right) combs with the same values of Δk as in Fig. 4. Resolution bandwidth (RBW) of the RF analyser was set to 10 kHz. One division of the vertical axis corresponds to 10 dB.

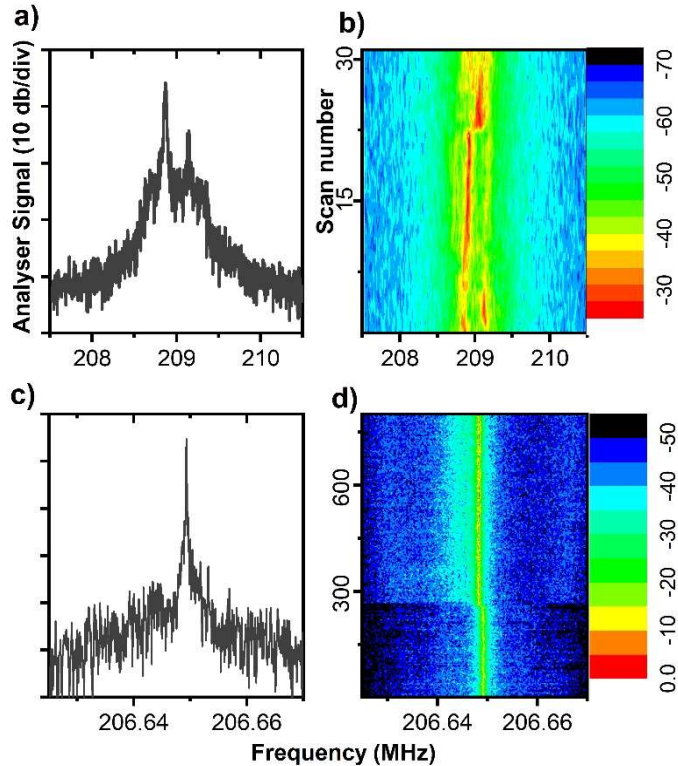

**Fig. 6**. (a) An RF spectrum of the fundamental comb (1615 nm) that shows a set of clearly different beat frequencies of the different subcombs. Measured with $\Delta k \sim -3.2$ mm$^{-1}$ (b) Evolution of the RF spectrum of (a) over several seconds. (The interval between subsequent scans is 0.4 s). RBW = 1 kHz. (c) Example of a narrow intermode beatnote measured with a setup that resonates the fundamental wavelength at 2036 nm; see Appendix for details. (d) Evolution of the RF spectrum of (c) over several minutes. RBW = 100 Hz. The sudden change was deliberatly induced, see text for details. Note the different frequency scales in different panels.

The most interesting operating regime is the one that corresponds to narrowing of the intermode beat signal to the kHz level or even below – which in the case of the setup reported here takes place in the vicinity of $\Delta k \sim -2.6$ mm$^{-1}$ (See Fig. 5). In time domain, the drastic narrowing of the beat note corresponds to pulsing of the optical field at a rate that corresponds to the mode-spacing frequency of ~209 MHz (Fig. A7 of the Appendix). In the best cases, the CW background disappears almost completely, similar to mode-locked lasers. Contrary to CQN mode-locked lasers, this state is however inherently unfavourable in a CQN cavity that lacks an energy reservoir. The pulsing inevitably reduces the overall nonlinear conversion efficiency, because the pump power can only be consumed during the pulses, and not between them. In our system, this is seen as a drop in the pump depletion, as measured from the 1064 nm pump beam of the initial singly-resonant OPO process that produces the fundamental beam (Fig. A4 of the Appendix). While the pump-depletion far from SHG phase matching exceeds 65 % even with the maximum pump power, the narrow-beat-note regime corresponds to a pump depletion of 40%. An example of a similar narrow-beatnote state for the 2036-nm setup is shown in Fig. 6c. A recording of the RF spectrum over five minutes is plotted in Fig. 6d, exemplifying that these states can be rather stable if not disturbed. The standard deviation of the beat frequency during a 15-minute measurement was just 400 Hz, even without any active stabilization of the cavity. The 400 Hz frequency jitter corresponds to about 3-µm variation in the cavity length. The sudden change of the beatnote frequency at scan number 258 was caused deliberately by slightly adjusting the cavity length by a piezoelectric actuator [17]. Despite the distinct changes in the RF spectra, the optical envelope spectra remain unchanged within the measurement precision of the optical spectrum analyser. It is also worth emphasizing that although the beat frequency stability can be relatively good over several minutes, the time-domain signal (Fig. A7 of the Appendix) typically exhibits notable amplitude variations in the time scale of ~ 1 second, even in the best case. The lack of long-term-stable pulse trains is also suggested by interferometric autocorrelation measurements that we have reported in our earlier work [5].

In some cases, the pump depletion drops to as low as 10 to 20 %. Also these regions often show narrow intermode beat signals, which however are very unstable and sensitive to even small variations of the cavity length. In general, we have observed that it is difficult to spontaneously reach and maintain long-term stable operation of the CQN comb in one of the low-noise regimes, and that any environmental disturbances or, *e.g.*, changes in the cavity length easily push the system to a state of high efficiency.

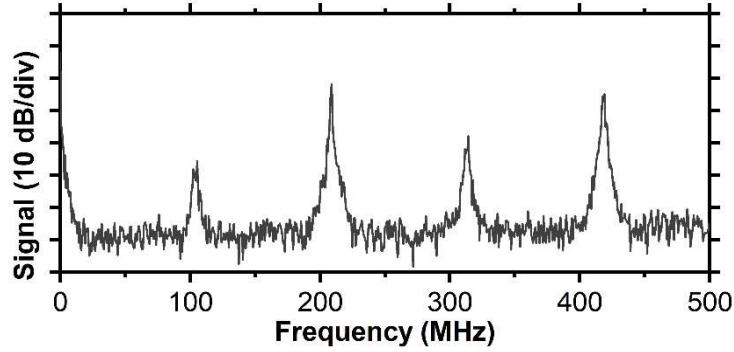

**Fig. 7**. RF spectrum of the fundamental comb at ~1615 nm, showing additional beat notes at half of the mode spacing, *i.e.* at ½ × 209 MHz. Measured with $\Delta k \sim -2.2$ mm$^{-1}$.

The problems with unstable mode-locking and low pump-to-comb conversion efficiency also exist with bright-soliton mode-locked $\chi^{(3)}$ Kerr combs, and it is likely that these problems can be solved using methods similar to those developed for Kerr combs. Along with proper cavity dispersion management, the abovementioned seeding method is a possible solution, but requires active modulation of the pump laser. The most common path into soliton states in Kerr combs uses "hard" excitation techniques, such as fast scanning of the pump laser frequency or "power kicking" [31]. Such methods should be directly applicable to CQN combs, at least with the microresonator platforms. In fact, we sometimes observe onset of the low-noise state by disturbing the cavity *e.g.* by knocking the optical table. Since our internally pumped system automatically adjusts such that the fundamental field resonates in the cavity, we cannot easily vary the cavity detuning, which is known to be an important parameter for both CQN comb [8] and Kerr comb [38] generation. The implementation of Fig. 1a would thus be better for a platform to experimentally investigate the effect of cavity detuning.

In general, dark-soliton operation would offer much better efficiency than bright-soliton mode-locking, and would thus be preferred for most applications [53, 54]. We have not observed any clear signs of dark-soliton mode-locking so far. Simulations may help identify experimental conditions that could favour such operation, for example in cases where group-velocity mismatch between the fundamental and second-harmonic frequencies is small [23, 55].

## 4. Conclusions and discussion

We have experimentally studied how the spectral characteristics of CW-pumped CQN combs depend on the phase mismatch $\Delta k$ of the first step of the CQN process, which in our case is second harmonic generation. The experiments were carried out using two different systems that were both internally pumped by a CW singly-resonant OPO. In addition to providing the resonant fundamental beam of the CQN comb, the CW SRO inherently converts the CQN comb to the mid-infrared region. Despite the differences in experimental details and fundamental wavelengths (1615 nm and 2036 nm), both systems show qualitatively similar behaviour in terms of phase mismatch. In particular, we have observed that it is possible to reach a regime of narrow intermode beat notes by adjusting $\Delta k$.

The values of $\Delta k$ were calculated from the PPLN refractive index data using the measured crystal temperature and poling period. These values are thus susceptible to experimental uncertainties, and should not be taken as absolute values. Rather, our observations reveal qualitative changes in the CQN comb characteristics as $\Delta k$ is varied. The calculated $\Delta k = 0$ corresponds to the observed maximum of SHG, which gives a one-point calibration of the $\Delta k$ axis and allows us to reliably determine the sign of $\Delta k$. The observation that the regimes of narrow intermode beat signals correspond to opposite signs of $\Delta k$ for the two setups (Fig. A4 of the Appendix) is likely to reflect the different dispersion properties of these two setups. In order to provide a benchmark for future theoretical work, we have included the experimental details in the Appendix.

We have treated the CW SRO simply as an intra-cavity pump source for the CQN, assuming that it does not otherwise influence the CQN comb generation process – except for converting the comb to the mid-infrared region. The main features of the observed CQN spectra can indeed be explained by a simple CQN model, which assumes SHG followed by parametric down conversion, independently of the CW SRO process. Future numerical modelling should take into account also other nonlinear effects. For instance, the parametric gain bandwidth of the SRO process limits the width of the mid-infrared comb, and is expected to influence the spectrum also in the fundamental spectral region via gain narrowing [44]. On the other hand, our SRO is operated multiple times above the threshold, which makes it susceptible to spectral broadening due to group-velocity-mismatch dependent modulation instability [56].

One of the remaining challenges with CQN combs is to achieve stable operation in a low-noise state. Prior numerical modelling and our experimental results suggest that such states can be accessed by soliton mode locking, similar to what has been reported for Kerr combs. Search of dark soliton states with nearly continuous output power and high conversion efficiency is a particularly interesting topic for future research.

Another (but related) remaining issue is to demonstrate full stabilization of the CQN comb. We have experimentally confirmed the CQN comb equidistance [19] and shown that it is possible to control the mid-infrared comb offset frequency and mode spacing independently [6]. However, the comb offset measurement and stabilization are yet to be implemented. The most common method, $f$-to-$2f$ interferometry, requires an octave-spanning spectrum. Such a significant spectral broadening – either inside or outside the comb generator – typically requires femtosecond pulses. Such pulses can potentially be produced afterwards by line-by-line shaping techniques even from a quasi-CW comb, if only the comb lines are mutually phase locked [52, 57].

## Acknowledgements

MV acknowledges the financial support of the Academy of Finland (314363). Both VU and MV acknowledge the support of the Flagship of Photonics Research and Innovation of the Academy of Finland.

# Appendix

## Experimental parameters

<u>Setup 1 (Fundamental wavelength 1615 nm)</u>

- CW-SRO pump laser system: IPG Photonic YAR-20K-1064-LP-SF amplifier seeded by a distributed feedback diode laser at 1063.5 nm (linewidth < 1MHz). Power at CW-SRO crystal 19 W in all measurements.
- CW-SRO crystal, 50 mm long periodically poled congruent MgO-doped $LiNbO_3$ (HC Photonics). Crystal temperature (T) controlled with a precision of 10 mK. In all measurements: Poling period 31.0 μm and T = 60 °C, leading to a CW SRO signal wavelength of 1615 nm. Focusing parameters of the pump, signal and idler beams in the crystal: $\xi$ = 2.
- CQN crystal, 50 mm long periodically poled congruent MgO-doped $LiNbO_3$ (HC Photonics). The poling period is adjustable by crystal translation between 19.5 and 21.3 μm in 0.3 μm steps. Crystal temperature (T) controlled with a precision of 10 mK. Focusing parameter of the signal (fundamental) beam in the crystal: $\xi$ = 2.
- Cavity mirrors: 2×PL (angle of incidence AOI = 20°) and 4×CC mirrors (AOI = 10°). Radius of curvature of the CC mirrors: -100 mm. All with a coating that is highly reflective at 1500 to 2300 nm, coated by Layertec. Reflectivity and group-velocity dispersion (GDD) plotted in Fig. A1 below.

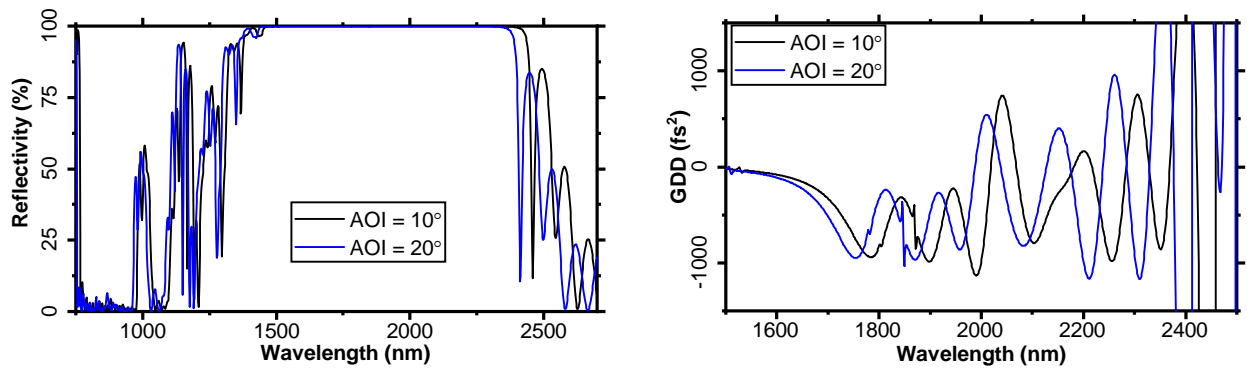

Fig. A1: Reflectivity and GDD plots of all cavity mirrors of Setup 1. For concave (CC) mirrors, angle of incidence (AOI) of 10° was used. The AOI at plane (PL) mirrors was 20°.

<u>Setup 2 (Fundamental wavelength 2030 nm)</u>

- Same setup as in V. Ulvila, C. R. Phillips, L. Halonen, and M. Vainio, "Spectral characterization of a frequency comb based on cascaded quadratic nonlinearities inside an optical parametric oscillator," Physical Review A, 92, p. 033816 (2015) http://link.aps.org/doi/10.1103/PhysRevA.92.033816.
- CW-SRO pump laser system: IPG Photonic YAR-20K-1064-LP-SF amplifier seeded by a distributed feedback diode laser at 1063.5 nm (linewidth < 1MHz). Power at CW-SRO crystal 19 W in all measurements.
- CW-SRO crystal, 50 mm long periodically poled (fan out) congruent MgO-doped $LiNbO_3$ (HC Photonics). Crystal temperature (T) controlled with a precision of 10 mK. In all measurements: Poling period 32.2 μm and T = 60 °C, leading to a CW SRO signal wavelength of 2036 nm. Focusing parameters of the pump, signal and idler beams in the crystal: $\xi$ = 2.
- CQN crystal, 50 mm long periodically poled ~~(fan out)~~ congruent MgO-doped $LiNbO_3$ (HC Photonics). The poling period is continuously adjustable by crystal translation between 26.5 and 32.5 μm (fan-

- out structure). Crystal temperature (T) controlled with a precision of 10 mK. In all measurements, T = 60 °C. The value of Δk varied by changing the poling period, which is adjustable between .26.5 and 32.5 μm. Focusing parameter of the signal (fundamental) beam in the crystal: ξ = 2.
- Cavity mirrors: 2×PL highly reflective at 1770 to 2050 nm (Layertec, reflectivity and group-velocity dispersion (GDD) plotted in Fig. A2 below). 4×CC mirrors similar to those used in Setup 1, highly reflective at 1500 to 2300 nm (see Fig. A1). ). Radius of curvature of the CC mirrors: -100 mm.

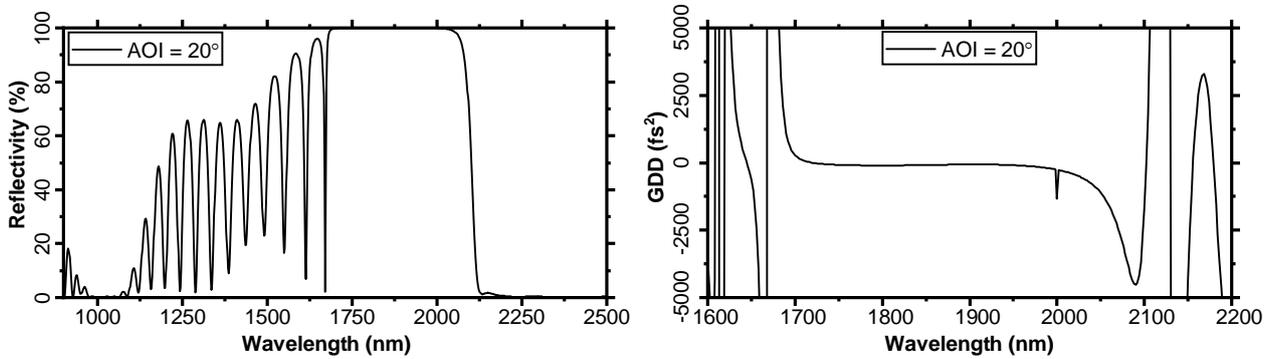

Fig. A2: Reflectivity and GDD plots of the plane mirrors of Setup 2. (The concave mirrors were the same as in Setup 1)

# Additional information

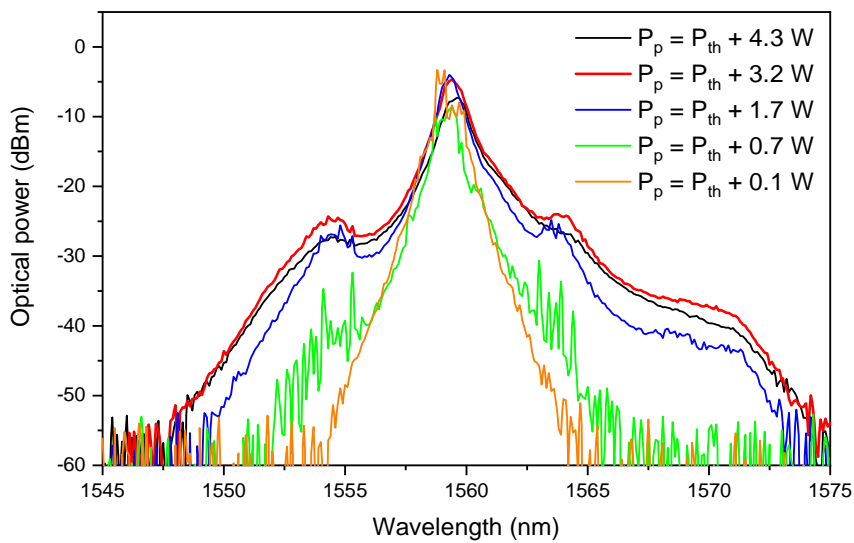

Fig. A3: The effect of pump power on the central peak of the fundamental spectrum. The spectra were recorded with a setup [1, 2] similar to Setup 1 that is described above, but with different set of cavity mirrors and at a different wavelength – despite of these differences in experimental parameters, the CQN comb properties are similar to those of Setups 1 and 2.

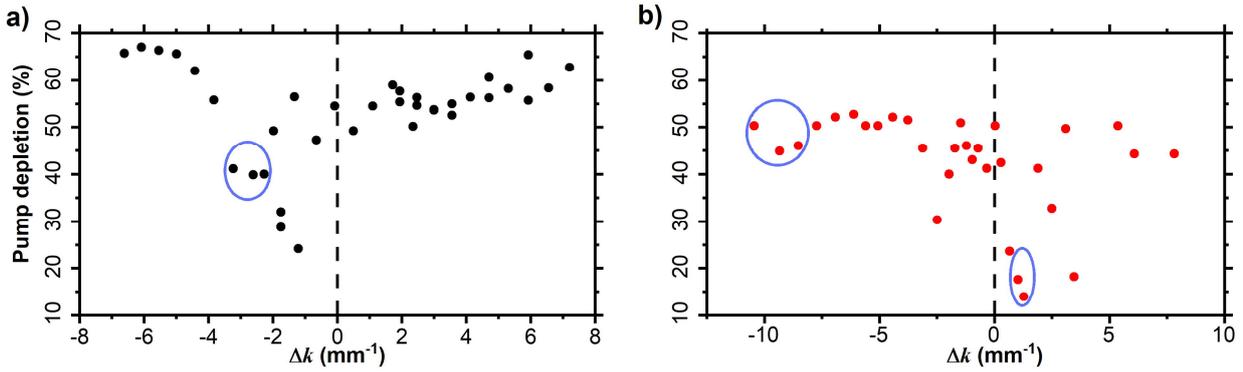

Fig. A4: The effect of Δk on CW-SRO pump depletion at 1063.5 nm. a) Setup 1 (fundamental wavelength 1615 nm). b) Setup 2 (fundamental wavelength 2036 nm). The circles indicate relatively stable regimes of narrow intermode beat signals. Note that, also in these regions, the system is typically sensitive to small changes of the cavity length.

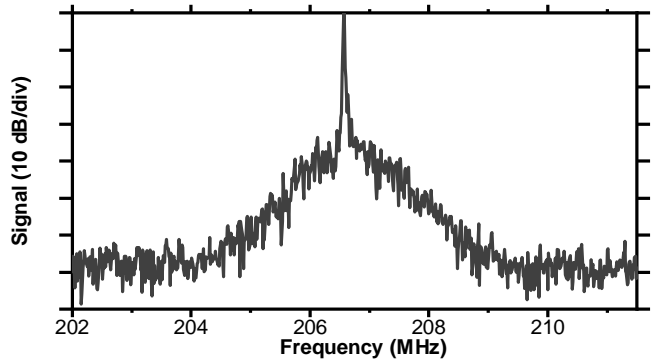

Fig. A5: Example of an RF spectrum recorded with Setup 2 in the regime of narrow (~ 1 kHz) intermode beatnote. The RF spectrum was recorded from the second harmonic spectrum of the fundamental (2036 nm) with Δk = +1.3 mm$^{-1}$ (QCN crystal: Λ = 30.94 μm, T = 60 °C) using the setup described in [3].

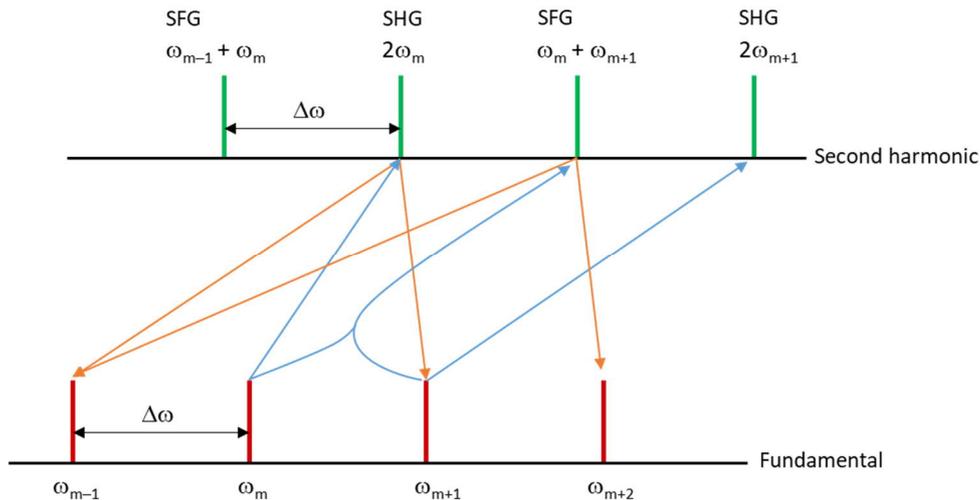

Fig. A6: Illustration of processes that can mutually couple and injection lock a CQN comb that has mode spacing Δω. The blue arrows show examples of possible up-conversion processes (SHG and SFG), and the orange arrows indicate parametric down conversion (DRO). As an example, the second harmonic ($2\omega_m$) of the fundamental comb mode $\omega_m$ can pump a DRO that produces signal and idler frequencies at $\omega_{m-1}$ and $\omega_{m+1}$ thus locking the phases of $\omega_{m-1}$, $\omega_m$ and $\omega_{m+1}$.

The same $2\omega_m$ mode can, within the parametric gain bandwidth, pump also other DRO processes that produce signal-idler pairs $\omega_{m-x}$ & $\omega_{m+x}$ where x is an integer.

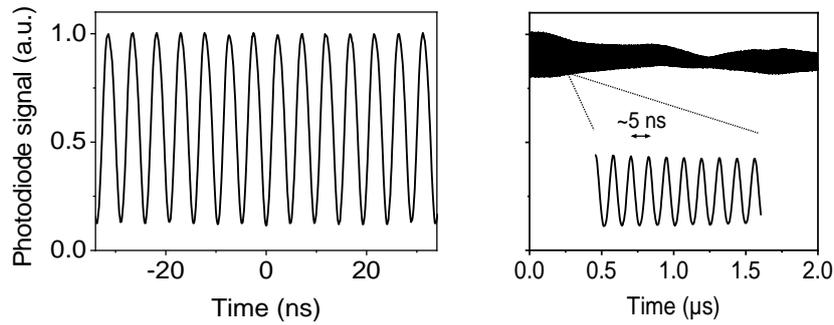

Fig. A7. Examples of time-domain signals measured in two regimes: Narrow beatnote (~1 kHz, left) and broad beatnote (~1 MHz, right). The oscilloscope used in the measurement limits the electrical measurement bandwidth to < 1 GHz.